\documentclass[conference]{IEEEtran}
\IEEEoverridecommandlockouts
\usepackage{cite}
\usepackage{amsmath,amssymb,amsfonts}
\usepackage{graphicx}
\usepackage{textcomp}
\usepackage{xcolor}

\usepackage[bookmarks,colorlinks]{hyperref}
\usepackage[linesnumbered,ruled,lined]{algorithm2e}
\usepackage{enumitem}
\usepackage{algpseudocode}
\usepackage{cite}
\usepackage{amsmath,amssymb,amsfonts,steinmetz}
\usepackage{graphicx}
\usepackage{mathrsfs}  
\usepackage{textcomp}
\usepackage{acronym}
\usepackage{xcolor}
\usepackage{times}
\usepackage{bm}
\usepackage{amsmath}
\usepackage{amssymb}
\usepackage{stmaryrd}
\usepackage{babel}
\usepackage{graphics, graphicx}
\usepackage{xcolor}
\usepackage{gensymb}
\usepackage{cite}
\usepackage{enumitem}
\usepackage{url}

\def\BibTeX{{\rm B\kern-.05em{\sc i\kern-.025em b}\kern-.08em
    T\kern-.1667em\lower.7ex\hbox{E}\kern-.125emX}}
\begin{document}

\title{A self-adaptive RIS that estimates and shapes \\ fading rich-scattering wireless channels}

\author{\IEEEauthorblockN{Chloé Saigre-Tardif}
\IEEEauthorblockA{\textit{Univ Rennes, CNRS, IETR - UMR 6164} \\
Rennes, France \\
chloe.saigre-tardif@insa-rennes.fr}
\and
\IEEEauthorblockN{Philipp del Hougne}\thanks{\textit{Corresponding Author: Philipp del Hougne.}}
\IEEEauthorblockA{\textit{Univ Rennes, CNRS, IETR - UMR 6164} \\
Rennes, France \\
philipp.del-hougne@univ-rennes1.fr}
}

\maketitle

\begin{abstract}
We present a framework for operating a self-adaptive RIS inside a \textit{fading rich-scattering} wireless environment. We model the rich-scattering wireless channel as being double-parametrized by (i) the RIS, and (ii) dynamic perturbers (moving objects, etc.). Within each coherence time, first, the self-adaptive RIS estimates the status of the dynamic perturbers (e.g., the perturbers' orientations and locations) based on measurements with an auxiliary wireless channel. Then, second, using a learned surrogate forward model of the mapping from RIS configuration and perturber status to wireless channel, an optimized RIS configuration to achieve a desired functionality is obtained. We demonstrate our technique using a physics-based end-to-end model of RIS-parametrized communication with adjustable fading (PhysFad) for the example objective of maximizing the received signal strength indicator. Our results present a route toward convergence of RIS-empowered localization and sensing with RIS-empowered channel shaping beyond the simple case of operation in free space without fading.

\end{abstract}


\section{Introduction}

We introduce an artificial intelligence (AI) empowered approach to converging channel estimation and shaping enabled by reconfigurable intelligent surfaces (RISs) in \textit{fading} rich-scattering wireless environments.
The ability of RISs to shape wireless channels within the emerging ``smart radio environment'' paradigm pivotally relies on channel state knowledge. Most algorithmic explorations of RISs assume that the channels are known, but gaining this channel knowledge once per coherence time in practice will involve significant overhead~\cite{zappone2020overhead}, a realization that has even given rise to ideas for how to use RIS without deterministic channel information~\cite{abrardo2021intelligent,albanese2021marisa,Kyriakos}. 

If the receiver is cooperative, the channels can be measured via pilot signals, as recently studied for RISs in free space~\cite{wang2020channel,liu2020matrix,hu2021two}. Alternatively, and this is the case we focus on, the RIS can be equipped with a sensing functionality to determine itself the channel state information. First efforts to integrate sensing functionalities for channel estimation into RISs have been reported for operation in free space where the problem of channel estimation is comparatively simple, requiring, for instance, only a direction-of-arrival estimation~\cite{ma2019smart,ma2020smart,alexandropoulos2021hybrid,alamzadeh2021reconfigurable,dai2022simultaneous}. But realistic propagation environments often involve complex fading effects significantly beyond the validity of a simple free-space approximation. Deployment scenarios of RISs inside metallic enclosures such as vessels, planes, trains or busses will give rise to channels with strong multipath characteristics; at low frequencies ($<$~6~GHz), even office rooms present substantial amounts of reverberation~\cite{Kaina_metasurfaces_2014,del2019optimally}. In such rich-scattering scenarios, the wireless channels are the superposition of many reflected waves arriving from seemingly arbitrary angles such that common free-space intuition and approaches cannot be applied~\cite{alexandropoulos2021reconfigurable}. The complexity of the problem calls for data-driven AI approaches.

\begin{figure*}[t]
    \centering
    \includegraphics[width=2\columnwidth]{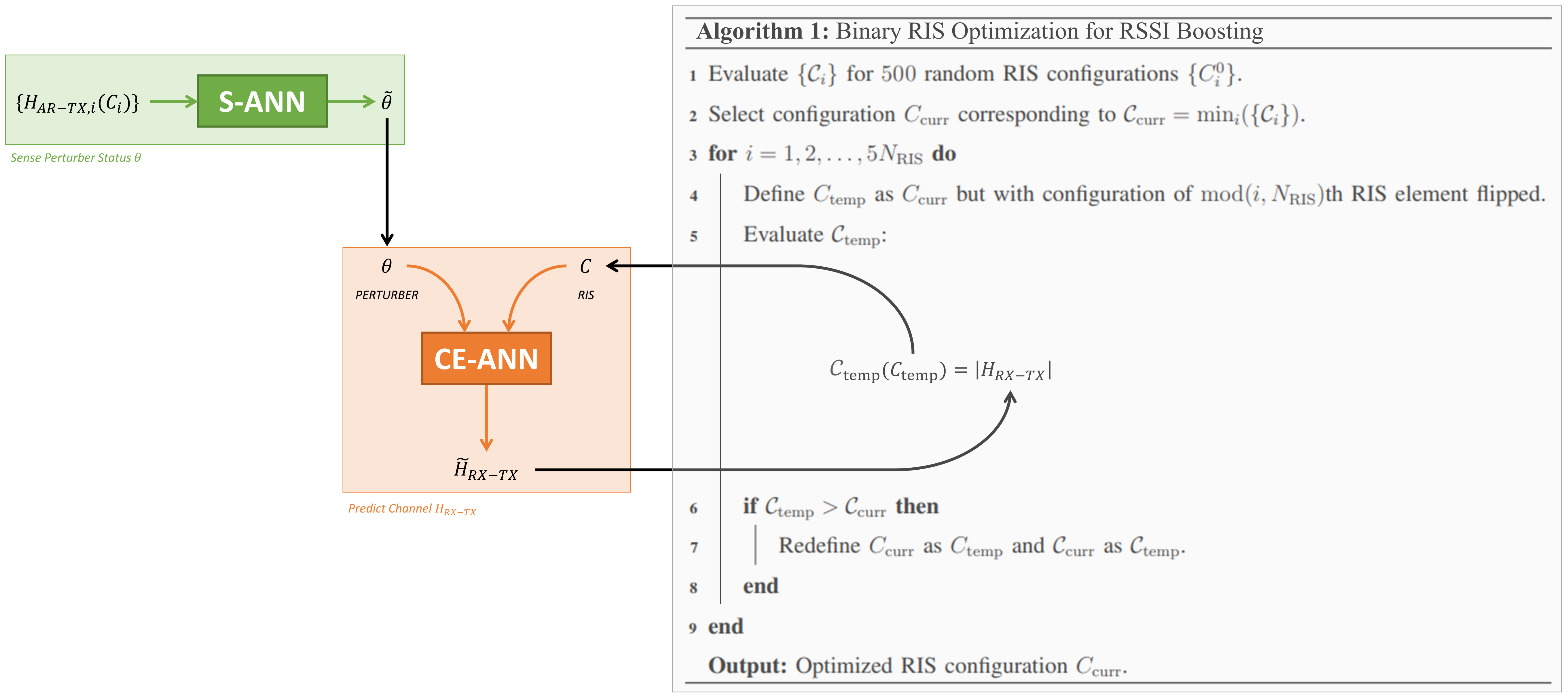}
    \caption{Algorithmic summary. First, the pertuber status $\theta$ is determined using S-ANN based on measurements of the transmitted spectrum between the transmitter (TX) and an auxiliary receiver (AR) for a fixed series of random RIS configurations. Second, a simple iterative algorithm (Algorithm 1) is used to optimize a RIS configuration for RSSI boosting. At every iteration, the cost function evaluation is based on the prediction by CE-ANN of the transmitted spectrum between TX and receiver RX. CE-ANN takes the estimated perturber status from S-ANN and the RIS configuration to be evaluated from Algorithm 1 as input.}
    \label{fig:concept}
\end{figure*}

In our approach, we consider a \textit{double-parametrization of the fading rich-scattering wireless channels}: 
\begin{enumerate}
  \item the desired RIS-parametrization.
  \item the undesired channel dependence on some dynamic channel-perturbing parameters such as the location of the receiver, the location of moving objects, the orientation of rotating objects, etc.
\end{enumerate}
We will thus proceed in three steps:
\begin{enumerate}
  \item Learn a surrogate forward model to predict the channel as a function of RIS parameters and perturbing parameters.
  \item Estimate the perturbing parameters.
  \item Optimize the RIS parameters, given the perturbing parameters, to achieve the desired shaping of the wireless channel.
\end{enumerate}
While step 1 is a one-off calibration step for a given scenario, steps 2 and 3 are repeated once within every channel coherence time.

The first step of learning a parametrized forward model of complex fading channels cannot be approached analytically in our rich-scattering setting (unlike its free-space counterpart). 

The second step can build on existing literature for RIS-assisted sensing in rich-scattering environments. Specifically, a fixed series of random RIS configurations can offer sufficient configurational diversity to estimate the location of a non-cooperative object based on transmission measurements between a single pair of nodes~\cite{del2018precise}, even deeply sub-wavelength precision with intensity-only measurements is attainable~\cite{del2021deeply}. This is essentially an instance of compressed sensing, where the rich-scattering environment acts as RIS-parametrized coded aperture. Importantly, the nodes can be positioned at any fixed but arbitrary point within the environment. This wave-fingerprinting-based localization technique can be implemented robustly in dynamic environments~\cite{LocalizationDynamicEnvironment} and it may be possible to learn a shorter task-specific fixed series of RIS configurations to improve latency through ``learned sensing''~\cite{del2021ris}. The related sensing task of object recognition (instead of localization) inside a rich scattering environment has also been implemented with compressed sensing using spectral diversity (broadband measurements)~\cite{frazier2021deep}.

The third step of optimizing the RIS parameters \textit{cannot} be conveniently implemented by running the forward model from step 1 in reverse with fixed weights and trainable inputs. The reason is the 1-bit (or few-bit) programmability of the RIS elements and the fact that the learned forward model is not guaranteed to interpolate in a physically meaningful way between the discretized inputs from the training data. 

Directly learning an inverse (instead of forward) model that outputs a suitable RIS configuration as a function of the desired wireless channel and perturber status does \textit{not} present itself as viable alternative to following steps 1 and 3 because (i) the forward model is a many-to-one mapping (within reasonable precision limits), and (ii) determining desired wireless channel characteristics in accordance with the laws of physics is nontrivial.

In the remainder of this paper, we detail our preliminary implementation of the above approach to realizing a self-adaptive RIS for channel sounding and shaping beyond the simple free-space case. To work with the substantially more challenging case of fading rich-scattering channels, we use an end-to-end model for RIS-parametrized fading channels that fully complies with wave physics~\cite{PhysFad}.

\section{Operation Principle}

The fading rich-scattering wireless channel between transmitter TX and receiver RX, $H_{RX-TX}$, is parametrized by (i) the RIS configuration, $C$, and (ii) the status of the dynamic perturber, $\theta$. $C$ is a $1\times N_{\textrm{RIS}}$ vector, where $N_{\textrm{RIS}}$ is the number of RIS elements. $\theta$ is a vector concatenating all necessary parameters to fully describe the pertuber status. In the simple case of a single irregularly shaped object rotating around its own axis, $\theta$ is simply the object's angular orientation. In general, however, the perturbation can depend on multiple parameters that would be collected in $\theta$. In addition, we assume the existence of an auxiliary wireless channel, $H_{AR-TX}$, between the transmitter and an auxiliary receiver, AR, located at a fixed arbitrary position.

An algorithmic summary of the self-adaptive RIS' workflow performed during each coherence time is provided in Fig.~\ref{fig:concept}. First, the self-adaptive RIS estimates the current state of the perturber, $\Tilde{\theta}$, by measuring the field transmitted from TX to AR, $H_{AR-TX}$, for a fixed series of random RIS configurations. To this end, the artificial neural network (ANN) coined S-ANN is used. Second, based on $\Tilde{\theta}$, the self-adaptive RIS identifies a suitable configuration in order to implement a desired channel shaping functionality. To that end, the RIS uses the simple iterative Algorithm~1; every evaluation of the cost function in Algorithm~1 makes use of a second ANN coined CE-ANN that estimates the channel between TX and RX for a given $\theta$ and $C$.

We use the ANNs (S-ANN and CE-ANN) as faithful interpolators of functions for which no analytical expression exists.

\textit{Remark 1:} For free-space operation, the channel $H_{RX-TX}$ is usually decomposed into the channel from the TX to the RIS, the RIS configuration, and the channel from the RIS to the RX. Such a simple linear cascaded model is not applicable in rich-scattering environments where each path may encounter different RIS elements along its complicated trajectory. The parametrization of the channel $H_{RX-TX}$ through RIS (and dynamic perturbers) is hence highly complicated and analytically intractable, which is why we learn the surrogate forward model CE-ANN.

\textit{Remark 2:} The stronger the perturber alters the auxiliary wireless channel $H_{AR-TX}$, the more accurately S-ANN can estimate $\theta$ with a limited sequence of measurements and/or in the presence of measurement noise~\cite{del2021deeply}. Various implementations of S-ANN (for different sensing tasks and measurement protocols) were reported in Refs.~\cite{del2018precise,LocalizationDynamicEnvironment,del2021deeply,alexandropoulos2021reconfigurable,del2021ris,frazier2021deep}.

\section{Problem Statement}

\begin{figure}[ht]
    \centering
    \includegraphics[width=1\columnwidth]{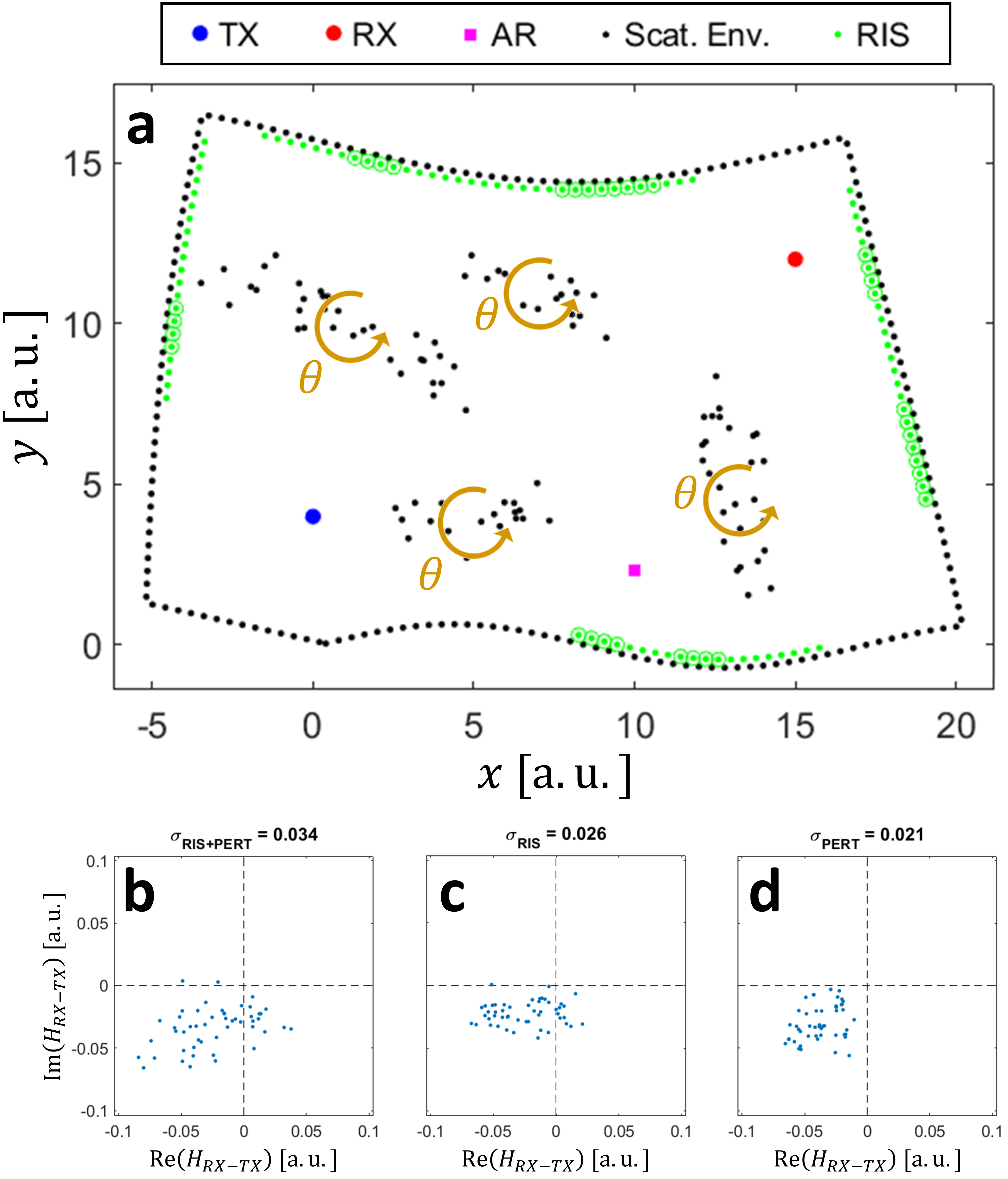}
    \caption{a) Considered setup in PhysFad~\cite{PhysFad}: An electrically large, irregularly shaped, rich-scattering enclosure inside which four irregular objects rotate (for simplicity in sync) in an uncontrolled manner. In addition, a distributed conformal RIS partially covers the enclosure's walls. The RIS consists of 100 1-bit programmable RIS elements which are clustered into macro-pixels of four RIS elements. An example configuration is indicated, where the RIS elements in the ON state are highlighted with a green circle. The wireless channel of interest, $H_{RX-TX}$, is parametrized by the RIS configuration $C$ and the perturber status $\theta$. The self-adaptive RIS is equipped with an auxiliary receiver (AR) at a fixed arbitrary location such that it can estimate the perturber status based on measurements of the auxiliary wireless channel $H_{AR-TX}$. b)-d) Distribution of $H_{RX-TX}$ in the complex plane for 50 random realizations of $C$ and $\theta$ (b), of $C$ with fixed $\theta$ (c), and of $\theta$ with fixed $C$. $\sigma$ denotes the standard deviation of the complex-valued $H_{RX-TX}$.
    }
    \label{fig:setup}
\end{figure}

For concreteness, we illustrate our proposed generic technique for the specific objective of maximizing the received signal strength indicator (RSSI, i.e., the magnitude of the transmitted spectrum) at a desired frequency between a fixed transmitter and a fixed receiver inside a rich-scattering enclosure including dynamic uncontrolled perturbers and a conformal distributed 1-bit-programmable RIS. As illustrated in Fig.~\ref{fig:setup}a, the perturbers are four irregularly shaped objects rotating in an uncontrolled manner around their centers. We model the wireless channels with PhysFad~\cite{PhysFad}, a physics-based end-to-end communication model for RIS parametrized wireless environments with adjustable fading. PhysFad is based on a coupled-dipole formalism constructed from first principles and thereby inherently obeys all essential physical principles, including

\begin{itemize}
 \item a notion of space and causality,
 \item dispersion (frequency selectivity) and the intertwinement of a RIS element's phase and amplitude response,
 \item any arising mutual coupling effects, including long-range mesoscopic correlations, and
 \item the \textit{non-linear} parametrization of wireless channels through RIS and perturbing objects.
\end{itemize}

PhysFad is formulated, without loss of generality, in arbitrary units
such that the central operating frequency as well as the medium’s permittivity and permeability are all defined to be unity~\cite{PhysFad}. The utilized PhysFad parameters are summarized in Table~\ref{tbl:PhysFadParameters}.

\begin{table} [h]
\caption{Summary of utilized PhysFad~\cite{PhysFad} parameters.}
\vspace{-0.4cm}
\label{tbl:PhysFadParameters}
\begin{center}
\begin{tabular}{ |c|c|c|c| } 
\hline
Entity & $f_\mathrm{res}$ & $\chi$ & $\Gamma^L$  \\
\hline
Transceivers & 1 & 0.5 & 0 \\ 
Scat. Env. & 10 & 50 & $10^4$ \\ 
RIS & $\{1,5\}$ & 0.2 & 0.05\\
\hline
\end{tabular}
\end{center}
\end{table}

The impact of the RIS and/or perturbers on $H_{RX-TX}$ is illustrated in Figs.~\ref{fig:setup}b-d. The standard deviation $\sigma$ of the distribution of the complex-valued $H_{RX-TX}$ for random choices of $C$ with fixed $\theta$ (Fig.~\ref{fig:setup}c) is seen to be comparable to that for random choices of $\theta$ with fixed $C$ (Fig.~\ref{fig:setup}d). If both $C$ and $\theta$ are chosen randomly, $\sigma$ is of course higher (Fig.~\ref{fig:setup}b). Overall, the non-trivial parametric dependence of $H_{RX-TX}$ on $C$ and $\theta$ is obvious, excluding the use of analytical channel models as in free space.

\section{Methods}
\label{sec:methods}

In this section, we present preliminary results for the three-step procedure outlined in the Introduction to realize a self-adaptive RIS under fading rich-scattering conditions. We leave discussions of technical details like the influence of the amount of training data and the choice of ANN hyperparameters for future work. We work on purpose with ``simple'' fully connected layers because the rich-scattering enclosure scrambles information such that the features we seek to identify are likely to be encoded in long-range correlations~\cite{LocalizationDynamicEnvironment,del2021deeply}.

\subsection{Step~1: CE-ANN}  

Our CE-ANN consists of two fully connected layers with 64 neurons and ReLu activation, followed by a linear output layer. Tackling a regression (rather than classification) problem, our cost function is the mean-squared-error (MSE) of the predicted $H_{RX-TX}$. For training, we use a labeled dataset with $10^5$ entries $\{C,\theta,H_{RX-TX}\}$ of which $15\%$ are used for validation to decide when to stop the training in order to avoid overfitting.

\subsection{Step~2: S-ANN}  

Our S-ANN consists of three fully connected layers with 256, 128 and 26 neurons, respectively, and ReLu activation, followed by a linear output layer. Tackling another regression problem, our cost function is again a MSE, this time the MSE of the predicted $\theta$ (defined in terms of arc length on a unit circle to account for the periodicity). For training, we use a labeled dataset with $10^3$ entries $\{\theta,\{H_{AR-TX,i}\}\}$ where the set $\{H_{AR-TX,i}\}$ denotes the collection of transmission spectra of the auxiliary wireless channel for a fixed series of 10 random RIS configurations. $15\%$ of the data are used for validation to decide when to stop the training in order to avoid overfitting.

\subsection{Step~3: Algorithm~1}  
Algorithm~1 is implemented as detailed in Fig.~\ref{fig:concept}, making use of CE-ANN everytime it evaluates the cost function. The $\theta$-input of CE-ANN is obtained from the output of S-ANN.

\section{Results}

\begin{figure}[t]
    \centering
    \includegraphics[width=1\columnwidth]{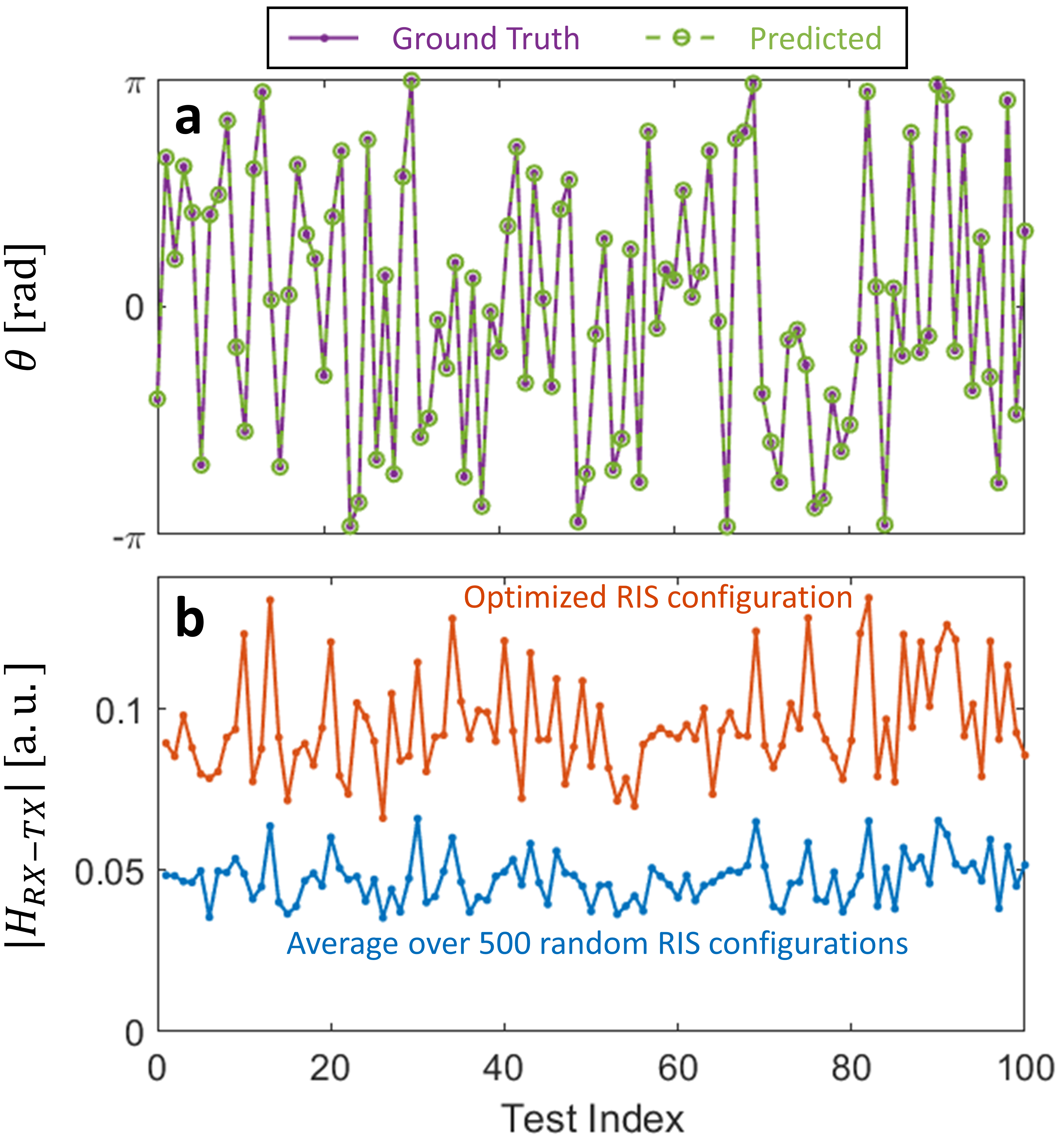}
    \caption{Results obtained for 100 test instances with random perturber status $\theta$. 
    a) Estimation of perturber status $\Tilde{\theta}$ (green) from S-ANN, and ground truth $\theta$ (violet).
    b) For each test instance, the average of RSSI over 500 random RIS configurations (blue) is contrasted with the RSSI of the RIS configuration optimized for the test instance using Algorithm~1 (red). While the identification of the optimized RIS configuration is solely based on CE-ANN, the final displyed RSSI results are evaluated in PhysFad~\cite{PhysFad} to avoid reliance on CE-ANN being a high-quality representation of the underlying model.
    }
    \label{fig:results}
\end{figure}

Our preliminary results are summarized in Fig.~\ref{fig:results}. After training S-ANN and CE-ANN as detailed in Sec.~\ref{sec:methods}, we test the performance of our self-adaptive RIS for the designated task of optimizing the RSSI on the wireless link $H_{RX-TX}$ for 100 random test instances of the dynamic rich-scattering environment. For each test instance, a random perturber status $\theta$ is chosen. The 10 measurements of the auxiliary channel $H_{AR-TX}$ for the 10 fixed random RIS configurations are made, and an estimate $\Tilde{\theta}$ of the perturber status is obtained. The results in Fig.~\ref{fig:results}a reveal excellent precision in estimating the perturber status. As noted above, estimating the perturber status is relatively easy for large perturbers as in the considered case.

Having obtained $\Tilde{\theta}$ in a given test instance, we next run Algorithm 1 to optimize the RIS configuration for the designated channel shaping functionality (RSSI maximization). The final optimized RIS configuration is evaluated in PhysFad to determine the achieved RSSI improvement independently from the quality of CE-ANN as a surrogate forward model. We also evaluate the RSSI for 500 random RIS configurations as benchmark. The results displayed in Fig.~\ref{fig:results}b indicate clear RSSI improvements in every test instance for our self-adaptive RIS. On average, the magnitude of $H_{RX-TX}$ is doubled, corresponding to a four-fold intensity improvement. 

\section{Conclusion}

We have introduced a paradigm through which a RIS can self-adapt its configuration inside a dynamic rich-scattering environment to reliably implement a desired channel shaping functionality despite dynamic perturbations. The conception of a self-adaptive RIS is particularly challenging to realize in our setting of a rich scattering environment where no simple analytical channel description exists. In practice, our technique will be useful when the channel coherence time is significantly longer than the time it takes our RIS to self-adapt through the procedure summarized in Fig.~\ref{fig:concept}. Current RIS prototypes can refresh their configuration within 20~$\mu$s (e.g., in Ref.~\cite{zhao2020metasurface}), with further room for improved refresh rates, suggesting that our method may serve in many realistic scenarios, e.g., where human motion is the perturbation. 

Looking forward, our technique can be extended to scenes with more complicated dynamics (e.g., complex motion of multiple perturbing objects and moving receiver). The impact of the SNR on the performance also deserves attention. Furthermore, we envision smarter ANNs that anticipate motion and cognitively select intelligent sensing patterns~\cite{saigre2021intelligent}. Overall, it remains to be determined whether the system level advantages of smart radio environments justify the cost of the associated overhead (in our work, going through the procedure summarized in Fig.~\ref{fig:concept}).

\bibliographystyle{IEEEtran}


\end{document}